# Out-of-Plane Polarization from Spin Reflection Induces Field-Free Spin-Orbit Torque Switching in Structures with Canted NiO Interfacial Moments

Zhe Zhang,[1] Zhuoyi Li,[1] Yuzhe Chen,[1] Fangyuan Zhu,[2] Yu Yan,[1] Yao Li,[1] Liang He,[1] Jun Du,[3] Rong Zhang,[1] Jing Wu,[4] Xianyang Lu,[1, 5, a)] Yongbing Xu[1, 4, 5, a)]

[1]*Jiangsu Provincial Key Laboratory of Advanced Photonic and Electronic Materials, School of Electronic Science and Engineering, Nanjing University, Nanjing 210093, China*
[2]*Shanghai Synchrotron Radiation Facility, Shanghai Advanced Research Institute, Chinese Academy of Sciences, Shanghai 201204, China*
[3]*Department of Physics, Nanjing University, Nanjing 210093, China*
[4]*York-Nanjing International Joint Center in Spintronics, Department of Electronics and Physics, University of York, York YO10 5DD, of UK*
[5]*School of Integrated Circuits, Nanjing University, Suzhou 215163, China*

**Realizing deterministic current-induced spin-orbit torque (SOT) magnetization switching, especially in systems exhibiting perpendicular magnetic anisotropy (PMA), typically requires the application of a collinear in-plane field, posing a challenging problem. In this study, we successfully achieve field-free SOT switching in the CoFeB/MgO system. In a Ta/CoFeB/MgO/NiO/Ta structure, spin reflection at the NiO interface, characterized by noncollinear spin structures with canted magnetization, generates a spin current with an out-of-plane spin polarization $\sigma_z$. We confirm the contribution of $\sigma_z$ to the field-free SOT switching through measurements of the shift effect in the out-of-plane magnetization hysteresis loops under different currents. The incorporation of NiO as an antiferromagnetic insulator, mitigates the current shunting effect and ensures excellent thermal stability of the device. The sample with 0.8 nm MgO and 2 nm NiO demonstrates an impressive optimal switching ratio approaching 100% without an in-plane field. This breakthrough in the CoFeB/MgO system promises significant applications in spintronics, advancing us closer to realizing innovative technologies.**

a) Authors to whom correspondence should be addressed: xylu@nju.edu.cn and ybxu@nju.edu.cn

In conventional electronics, electron spin has been underutilized, limiting devices to the charge degree of freedom. Spintronics revolutionizes microelectronics by unlocking the potential of electron spin, providing additional degrees of freedom and enabling the construction of various devices[1-3]. Spin-orbit torque (SOT) emerges as a superior technique, surpassing traditional spin-transfer torque (STT) with reduced power consumption and faster operation. This approach involves inducing a transverse spin current at the bilayer interface, manipulating the magnetization of the ferromagnetic layer[4]. Notably, SOT shows promise in SOT-MRAM cells, offering improved design margins and stability compared to STT-MRAM schemes[5].

However, achieving deterministic current-induced SOT magnetization switching, particularly in systems with perpendicular magnetic anisotropy (PMA), presents challenges, often requiring a collinear in-plane field, which imposes significant constraints on device integration and scalability. Such limitations are not conducive to the design principles of high-density and thermally stable SOT devices. Recent advancements aim at field-free SOT switching in PMA devices, exploring various techniques like exchange bias induced by an adjacent antiferromagnet[6,7], stray fields from neighboring ferromagnetic layers[8], structure engineering[9,10] and designing for tilted magnetic anisotropy[11]. These approaches may be vulnerable to current-induced Joule heating, encounter constraints in material choices necessitating larger write currents, and may not be applicable to practical large wafer-scale fabrication processes. These challenges underscore the need for further advancements in the field.

In contrast, the utilization of out-of-plane spin polarization ($\sigma_z$) generated by peculiar crystal symmetries represents an efficient and more universally applicable approach[12][13]. $\sigma_z$ typically emerges in two-dimensional van der Waals semimetals with lower crystal symmetry, such as $WTe_2$ and $MoTe_2$[14,15]. Recent reports also indicate its generation in certain nonlinear antiferromagnetic materials like $Mn_3Sn$ and $Mn_3GaN$[16,17]. Controlling crystal symmetry and the resulting $\sigma_z$ remains highly challenging, and it is incompatible with integration into semiconductor-based complementary metal-oxide-semiconductor (CMOS) technology. Therefore, the generation of $\sigma_z$ conventional in CoFeB/MgO heterostructures for achieving field-free switching is crucial for practical spintronics applications, given the widespread use of CoFeB as the free layer in high-performance magnetic tunnel junctions[18].

In this work, we propose a Ta/CoFeB/MgO/NiO/Ta structure, which successfully achieves efficient field-free SOT switching by generating out-of-plane spin polarization. Additionally, the use of NiO as an antiferromagnetic insulator in the device prevents the current shunting effect, providing good thermal stability with a Néel temperature of 520 K[19]. The canted magnetization and internal electric field at the NiO interface cause the reflected spins to flip, rotate, and precess. This interaction ultimately reorients the spin polarization to include an out-of-plane component. To confirm the contribution of the $\sigma_z$ to the field-free SOT switching, we conducted measurements of the shift effect in the out-of-plane hysteresis loops under different currents. Spin reflection at the NiO interface generates a spin current with out-of-plane spin polarization $\sigma_z$, overcoming the limitation of external magnetic fields. As depicted in Fig. 1(b), the z-polarized spins induce an out-of-plane anti-damping torque M × z × M ($\tau_z$), enabling deterministic switching of the CoFeB layer without the necessity of an external magnetic field. The sample with 0.8 nm MgO and 2 nm NiO demonstrates an optimal switching ratio of up to 93% at an applied in-plane field $H_x$ = 0 Oe. As the NiO layer thickness increases, the canting of the spin direction of the interfacial NiO layer moments induced by spin-flop coupling diminishes. The ratio of field-free SOT switching gradually decreases with the increasing thickness of the NiO layer, and no field-free switching is observed when the NiO layer thickness reaches 10 nm. This achievement in the CoFeB/MgO system holds vast potential for applications in spintronics, including spin-orbit torque-logic devices, neuromorphic computing, and deep neural networks[1].

All the samples were grown on Si substrates with $SiO_2$ insulation layers using a magnetron sputtering system. The layer sequence Ta (5)/$Co_{40}Fe_{40}B_{20}$ (1)/MgO (n)/NiO (t)/Ta (2) was deposited from bottom to top, as shown in Fig. 1(a). The numbers in parentheses denote thickness in nanometers. In subsequent discussions, these samples are referred to as MgO (n)/NiO (t) samples with different film thicknesses. The NiO layers were grown by sputtering a pure Ni metal target in a magnetron sputtering system with a mixture of argon and oxygen gases[20]. During the film growth process, a 50 sccm Ar gas flow was introduced, while the flow rate of $O_2$ was 3 sccm. X-ray diffraction (XRD) pattern in Fig. 1(c) reveals a distinct NiO (111) peak for the MgO (0.8)/NiO (30) sample. Simultaneously, the Reflection High-Energy Electron Diffraction (RHEED) patterns indicate a polycrystalline nature of the NiO layer grown on the MgO (refer to Supplementary Material). The cross-sectional transmission electron microscope (TEM) image of the device structure shown in Fig. 1(d) reveals a high-quality multilayer structure, highlighting the continuity of the 0.8 nm MgO layer. The stack was then post-growth annealed at 300 ℃ for 0.5 h in vacuum. The static magnetic hysteresis loops of the MgO (0.8)/NiO (2) sample, measured at room temperature using vibrating sample magnetometer (VSM) and Magneto-optical Kerr effect (Moke) measurements, reveal a stable perpendicular magnetic anisotropy (PMA), as illustrated in Supplementary Material. Hall bar devices with a width of 10 μm were fabricated for these stacks through optical lithography and etching process. Fig. S3 presents an optical image of the Hall bar device and schematic illustration of the electric measurements. The direction perpendicular to the plane is defined as the z-axis, with pulse current applied along the x-axis, and lateral Hall voltage measured along the y-axis.

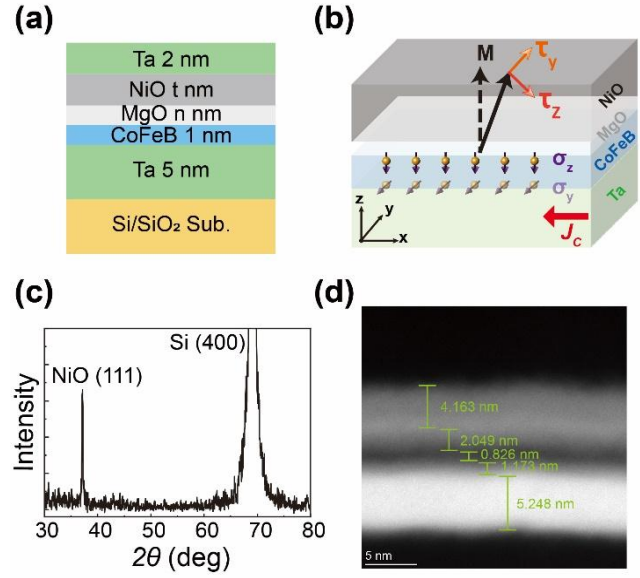

FIG 1. (a) Schematic representation of our samples. (b) Schematic of our SOT device. The z-polarized spins ($\sigma_z$) induce an out-of-plane anti-damping torque M × z × M ($\tau_z$), facilitating deterministic switching of the CoFeB layer without the need for an external magnetic field. (c) X-ray diffraction patterns of the MgO (0.8)/NiO (30) sample. (d) The high-resolution cross section transmission electron microscope (TEM) image of the multilayers.

The Anomalous Hall effect (AHE) loops, measured with an out-of-plane external magnetic field ($H_z$), are shown in Fig. 2(a) for MgO (2)/NiO (2), MgO (1)/NiO (2), MgO (0.8)/NiO (2) and MgO (0.8)/NiO (0) samples. It is evident that, even with a reduced MgO layer thickness of 0.8 nm, clear PMA is maintained[12,13,21,22]. Additionally, it is observed that the sample without the NiO layer exhibit a relatively larger coercive force. Subsequently, we applied pulsed direct current ($I_{pluse}$) and measured the change in Hall resistance for Hall bars while the in-plane auxiliary field ($H_x$) varying from + 20 Oe to - 20 Oe. The current pulse width was maintained at 100 μs. Following each pulse, the Hall voltage was measured by applying a small reading current of 200 µA. The switching curves of Hall resistance ($R_H$) are shown in Fig. 2(b, c, e and f) under varying $H_x$ from + 20 Oe to - 20 Oe. It is evident that the MgO (2)/NiO (2) and MgO (1)/NiO (2) samples exhibit significant SOT switching only when the $H_x$ exceeds approximately ± 8 Oe. Nevertheless, an almost complete SOT switching loop was observed at $H_x = 0$ Oe when the MgO layer thickness was reduced to 0.8 nm, as illustrated for the MgO (0.8)/NiO (2) sample in Fig. 2(e). Furthermore, almost no switching signal was detected in this sample when $H_x = - 5$ Oe. These findings suggest the existence of an effective internal field in this structure, contributing to the deterministic field-free SOT switching process[3,13]. Furthermore, we conducted MOKE image measurements after each pulse current at $H_x = 0$ Oe. The observed current-induced domain evolution is depicted in Fig. 2(d), offering compelling evidence for field-free switching in the MgO (0.8)/NiO (2) sample[23]. Based on the aforementioned results, field-free switching is exclusively observed when the MgO layer thickness is reduced to 0.8 nm in samples with a 2 nm NiO layer. As the MgO thickness further decreases, the multilayer film structure fails to establish stable PMA (refer to Supplementary Material). To investigate the impact of the NiO layer on magnetization switching, we prepared a comparative sample, MgO (0.8)/NiO (0). Remarkably, no indication of field-free switching was observed in the sample without a NiO layer, as shown in Fig. 2(f). These findings strongly support the crucial role of the NiO layer in enabling field-free switching.

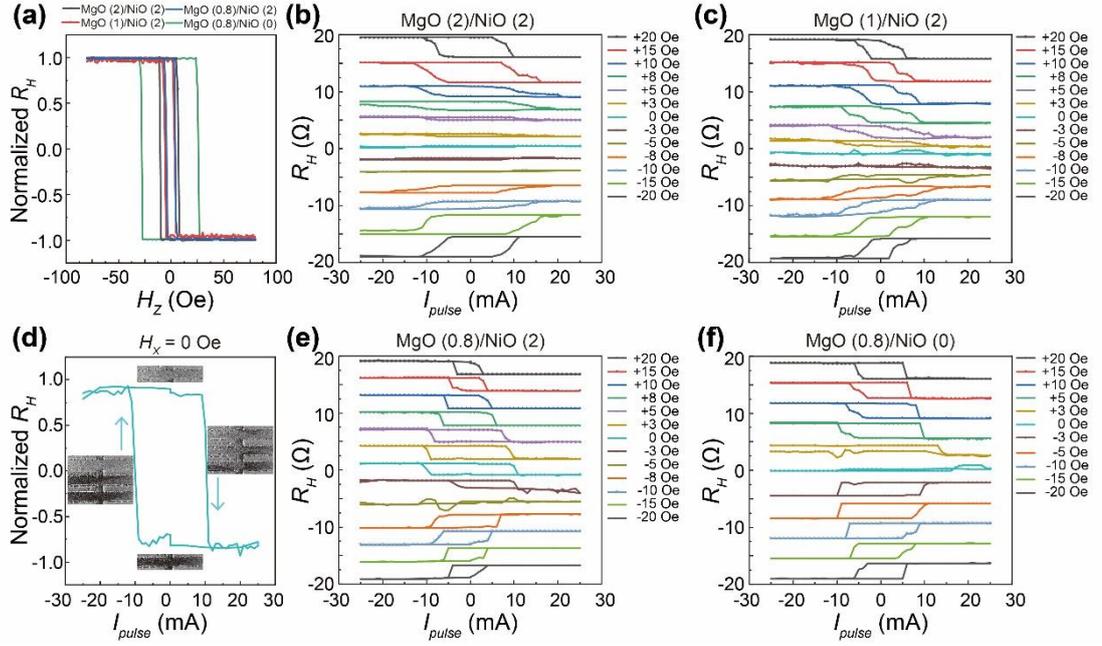

FIG 2. (a) Anomalous Hall effect loops for MgO (2)/NiO (2), MgO (1)/NiO (2), MgO (0.8)/NiO (2) and MgO (0.8)/NiO (0) samples. Current-induced magnetization switching under different magnetic field ($H_X$) for (b) MgO (2)/NiO (2), (c) MgO (1)/NiO (2), (e) MgO (0.8)/NiO (2) and (f) MgO (0.8)/NiO (0) samples. (d) Zero-field switching curve and MOKE images of magnetization switching for the MgO (0.8)/NiO (2) sample.

Next, we explore the mechanism underlying the field-free switching in MgO (0.8)/NiO (2) sample. As reported in previous studies, the antiferromagnetic insulator NiO may introduce exchange bias, obtaining an out-of-plane SOT effective field for field-free switching[6,7]. However, in our sample, we introduced a MgO layer between the CoFeB and NiO layers to mitigate the exchange bias effect. The hysteresis loops for the in-plane and out-of-plane directions, as shown in Fig. S2, also distinctly demonstrate the absence of exchange bias in the MgO (0.8)/NiO (2) sample. Another plausible mechanism for field-free switching involves the realignment of part of the spin current with a spin polarization $\sigma_y$, facilitated by the insertion of the NiO layer.

This realignment induces spin polarization with an out-of-plane component $\sigma_z$[17,24-28]. To validate the presence and contribution of the out-of-plane component $\sigma_z$ to the field-free SOT switching in our samples, we conducted measurements of the shift effect in the out-of-plane hysteresis loops under different currents[12-14,21,29]. Fig. 3(a, b) illustrate the AHE loops under ± 1 mA and ± 6 mA at an in-plane field $H_x$ = 0 Oe for the MgO (0.8)/NiO (2) sample, respectively. The distinct right and left shift of the AHE loop was observed when the current is ± 6 mA at $H_x$ = 0 Oe. The calculation of the shift field $|\Delta H|$ is determined in the inset of Fig. 3(b). The shift field $|\Delta H|$ at $H_x$ = 0 Oe with current for MgO (0.8)/NiO (2) and MgO (0.8)/NiO (0) samples is summarized in Fig. 3(e). For the MgO (0.8)/NiO (2) sample, with the current smaller than ± 1 mA, no obvious shift was detected. However, upon increasing the current to ± 2 mA, an abrupt shift was observed, and the value of $|\Delta H|$ continued to increases linearly with the escalating current, resembling findings previously reported with the presence of the spin current with an out-of-plane component $\sigma_z$[12,13,25]. For the MgO (0.8)/NiO (0) sample, it was observed that the shift field $|\Delta H|$ remained nearly zero at $H_x$ = 0 Oe, regardless of the magnitude of the current. Compared with the sample without NiO layer, the pronounced shift effect in the AHE loops at $H_x$ = 0 Oe for the MgO (0.8)/NiO (2) sample clearly confirms the existence of the out-of-plane component $\sigma_z$.

Fig. 3(f) presents the comparable results of $|\Delta H|$ for the MgO (0.8)/NiO (2) and MgO (0.8)/NiO (0) samples with $H_x$ = 500 Oe. The linear slope between $|\Delta H|$ and current can quantitatively manifest the SOT efficiency[30]. From the results, the MgO (0.8)/NiO (2) sample exhibits a higher SOT efficiency, with a slope of 11.28 ± 0.26 Oe/mA, whereas for the MgO (0.8)/NiO (0) sample, the slope is 7.84 ± 0.08 Oe/mA. Additionally, we conducted harmonic Hall voltage measurements by sweeping the in-plane magnetic field to further investigate the SOT efficiency[31-33]. Fig. 3(c, d) illustrate the typical first ($V_\omega$) and second ($V_{2\omega}$) harmonic voltage signals for the MgO (0.8)/NiO (2) sample under 1 mA (AC) with longitudinal field $H_x$ (H // I) and transverse field $H_y$ (H ⊥ I), respectively. The insets in Fig. 3(c, d) illustrate the fitting process of the data. Based on the Hall voltage results, longitudinal effective field $\Delta H_L$ and transverse effective field $\Delta H_T$ generated by the damping-like and field-like SOTs are extracted shown in Fig. 3(g, h). The values of $\Delta H_L$ and $\Delta H_T$ can be extracted by fitting the harmonic voltage curves under a small magnetic field range using the following equation[33]:

$$\Delta H_{L(T)} = -2 \frac{dV_{2\omega}/dH_{L(T)}}{d^2V_\omega/dH^2_{L(T)}}. \quad (1)$$

Both the SOT effective fields $\Delta H_L$ and $\Delta H_T$ are enhanced in the sample with a NiO interlayer. Specifically, the MgO (0.8)/NiO (2) sample exhibited a SOT effective fields of $\Delta H_L$ = 31.69 ± 1.85 and $\Delta H_T$ = 98.04 ± 2.26 Oe/mA, while the MgO (0.8)/NiO (0) sample showed $\Delta H_L$ = 27.27 ± 2.16 and $\Delta H_T$ = 78.18 ± 3.94 Oe/mA.

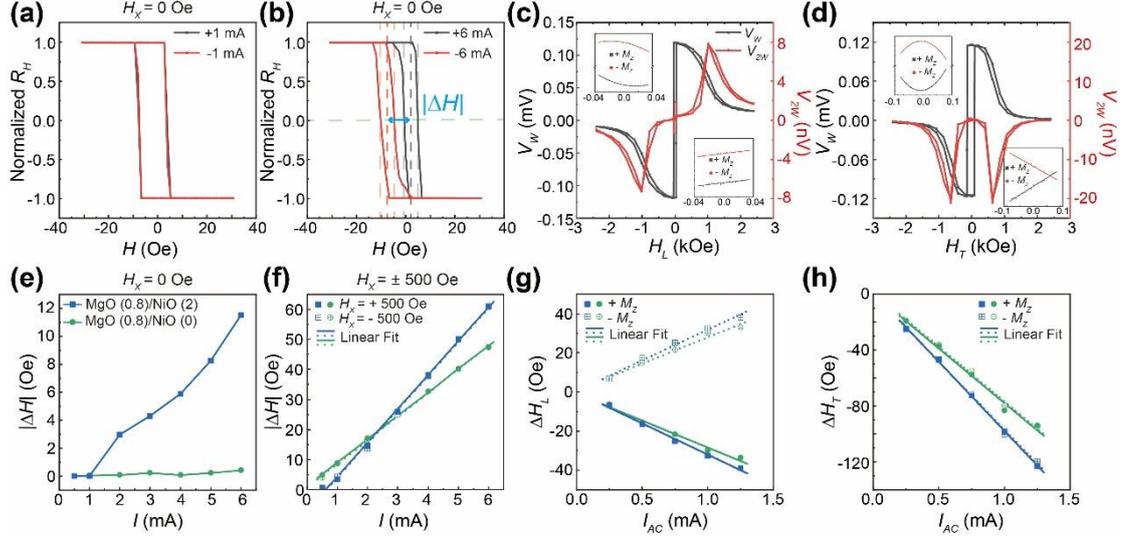

FIG 3. Shift effect of the out-of-plane hysteresis loops under (a) ± 1 mA and (b) ± 6 mA at an in-plane field $H_x$ = 0 Oe for MgO (0.8)/NiO (2) sample. The calculation of the shift field |Δ$H$| is determined in the inset of (b). (e, f) Summarized |Δ$H$| with current at $H_x$ = 0 and ± 500 Oe. First and second harmonic Hall signals as a function of the in-plane magnetic field (c) $H_x$ (H∥I) and (d) $H_y$ (H⊥I) under 1 mA. The current-induced effective fields versus $I_{AC}$ value for (g) longitudinal effective field Δ$H_L$ and (f) transverse effective field Δ$H_T$.

Above discussions demonstrate that the insertion of the NiO layer and the reduction in the thickness of the MgO layer in the multilayer structure lead to the generation of $\sigma_z$ in the system, thereby achieving field-free SOT switching. To delve deeper into the mechanism of $\sigma_z$ generation, we prepared a series of additional samples. Firstly, we can rule out the generation of spin currents by the Ta capping layer and its interaction with the NiO layer to produce $\sigma_z$. The 2 nm thick Ta layer, acting as the capping layer, will undergo complete oxidation to $Ta_2O_5$[34,35], acquiring insulating properties and rendering it no longer viable as a spin source. To provide further evidence, we replaced the capping layer material with Ru. The spin Hall angle of Ru has an opposite sign compared to Ta and is significantly smaller[36-38]. As shown in Fig. S5, the MgO (0.8)/NiO (2) sample with Ru capping also achieved field-free switching with a ratio approaching complete switching. The above results rule out the possibility that the spin polarization of the spin current has an out-of-plane component $\sigma_z$ originating from the capping layer Ta.

Subsequently, we investigated the influences of reflected spin on SOT switching, as depicted in Fig. 4(c). When electrons traverse the Ta layer in the x-direction, a spin current in the z-direction with polarization along the y-direction is induced. Due to the low intrinsic Gilbert damping parameter and the high effective spin-mixing conductance with moderately high transparency of the Ta/CoFeB bilayer system[39], this spin current is injected into the adjacent magnetic CoFeB layer, exerting torques on the magnetization of the CoFeB layer[1,40]. However, at this point, the spin current only exhibits $\sigma_y$, lacking the essential spin polarization component in the out-of-plane direction, $\sigma_z$, thus rendering it incapable of achieving field-free switching. Spin electrons polarized along the y-direction undergo spin reflection at the interface of NiO, thereby generating $\sigma_z$. The significant spin reflection can be attributed to the difference in work function between the two materials, resulting in the generation of an internal electric field pointing from NiO to CoFeB[13,21,41-43]. The spin current carrying an out-of-plane spin polarization component $\sigma_z$ is subsequently reflected back to the CoFeB layer, facilitating the field-free switching. This requires the MgO insulating tunnel barrier to have a sufficiently thin thickness to achieve a greater intensity of tunneling current while preserving the integrity of the internal electric field. This aligns with the experimental observations, where field-free switching was observed when the MgO layer thickness was reduced to 0.8 nm in the MgO (0.8)/ NiO (2) sample. This also suggests that the interfacial magnetic and electronic structure at the NiO/MgO interface is critical and should be thoroughly considered.

It is noteworthy that the existence of antiferromagnetic Néel vector in NiO at the interface is crucial. In our sample, the crystal orientation of NiO is (111). Under normal circumstances, the Néel vector of NiO aligns along the in-plane direction since the easy plane is along the (111) orientation. The interfacial magnetic structures were probed using X-ray Magnetic Circular Dichroism (XMCD) in Fig. 3(a, b). Intriguingly, the Ni $L_{3,2}$ edges exhibit XMCD signals in the MgO (0.8)/ NiO (2) sample, indicating a net magnetic moment arising from the Ni atoms. The values of spin moment $m_s$ = 0.390 $\mu_B$/atom and orbital moment $m_l$ = 0.067 $\mu_B$/atom were calculated by applying sum rules (refer to Supplementary Material) on the integrated XMCD and total XAS spectra. The spin moment is approximately 20% of that for bulk NiO crystals, as determined from magnetic x-ray scattering measurements[44,45]. However, no Ni $L_{3,2}$ edge XMCD signal was observed in the MgO (2)/NiO (2) sample. This demonstrates the clear role of ferromagnetism in inducing a ferromagnetic ordering of interfacial Ni spins, which coincides with the conditions necessary for the generation of $\sigma_z$. Fig. S7 shows The Ni $L_{2,3}$-edge x-ray absorption (XAS) spectra of the NiO in the MgO (0.8)/NiO (10), MgO (0.8)/NiO (2) and MgO (0.8)/Ni (2) samples. For a NiO thickness of 10 nm, the XAS characteristic spectra observed are consistent with those reported in the literature[46,47]. However, for the 2 nm NiO sample, the $L_2$ and $L_3$ peaks exhibit larger full widths at half maximum (FWHM) compared to the Ni thin film sample, and distinct peaks are not evident. This observation is consistent with report on thin NiO in the literature[45]. This suggests that thin NiO films may have certain defects, which in turn weaken their intrinsic antiferromagnetic exchange interactions. When the MgO layer is sufficiently thin, the NiO layer is influenced by interactions with adjacent ferromagnetic layers[46-49]. This results in a slight canting of the spin direction of the interfacial NiO layer moments away from the AFM in-plane easy axis[45]. In other words, the Ni moments at the interface tilt towards the Co moments due to the Co-Ni exchange interaction, leading to the formation of noncollinear spin structures at the interface of the NiO layer. The canted magnetization and the presence of an internal electric field at the NiO interface interact with the reflected spin, causing spin flip, rotation, and precession. This interaction ultimately reorients the spin polarization to include an out-of-plane component.

In addition to the MgO (0.8)/ NiO (2) sample, we also prepared samples with NiO thicknesses of 3, 5, and 10 nm to further validate our explanation. As shown in Fig. 4(d), in the current-induced magnetization switching experiments, the ratio of field-free SOT switching gradually decreases with the increasing thickness of the NiO layer, and no field-free switching is observed when the NiO layer thickness reaches 10 nm. This is due to the fact that the strength of this coupling depends on the competition between the exchange interaction of the interfacial magnetic moments in the antiferromagnetic and ferromagnetic layers, as well as the antiferromagnetic super-exchange interaction within the NiO layer[45]. Moreover, with increasing thickness of the NiO layer, enhanced effective AFM exchange coupling results in a reduction of the canting angle[50]. Consequently, this weakens the return of the spin current carrying an out-of-plane spin polarization component $\sigma_z$ through reflection. We experimentally confirmed this phenomenon by investigating the shift threshold effect of the AHE loops for MgO (0.8)/NiO (2, 3, 5 and 10) samples under the in-plane field $H_x = 0$ Oe, as depicted in Fig. 4(b). As the NiO thickness increases, the threshold current value gradually increases. At the same time, the shift field $|\Delta H|$ decreases accordingly. These results qualitatively indicate a reduction in the generation of spin current with an out-of-plane component $\sigma_z$, which is correlated with the ratio of field-free switching in these samples.

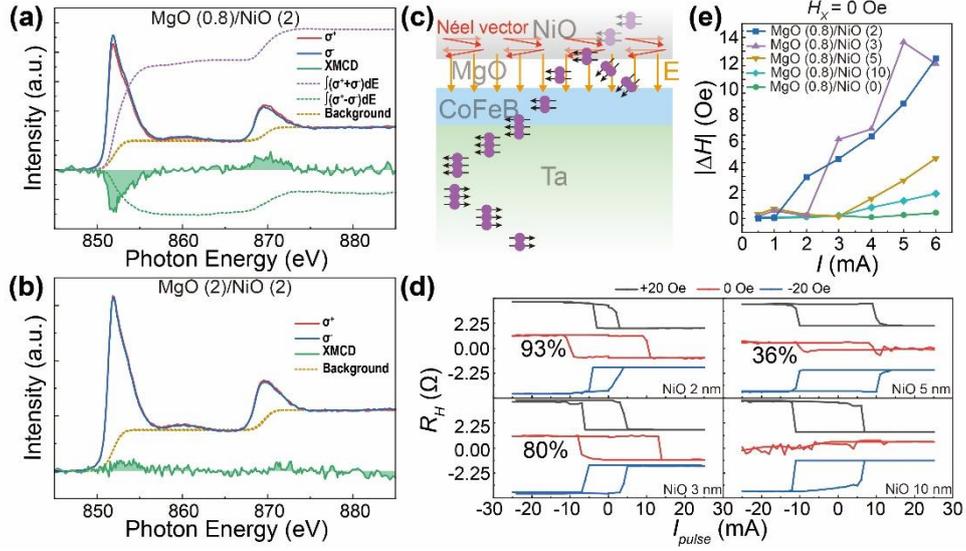

FIG 4. (a, b) Ni $L_{2,3}$-edge x-ray absorption (XAS) and x-ray magnetic circularly dichroism (XMCD) spectra of the NiO in the MgO (0.8)/NiO (2) and MgO (2)/NiO (2) samples. (c) Schematic representation of spin tunneling and spin reflection in the multilayer film structure. (d) Current-induced magnetization switching under different magnetic field ($H_X$) for the MgO (0.8)/NiO (t) samples (t = 2, 3, 5 and 10 nm). (e) Summarized shift field |$\Delta H$| with current at $H_x$ = 0 Oe for the MgO (0.8)/NiO (t) samples (t = 0, 2, 3, 5 and 10 nm).

In summary, we have successfully achieved field-free SOT switching within a Ta/CoFeB/MgO/NiO/Ta structure. Notably, the MgO (0.8)/NiO (2) sample exhibited an impressive switching ratio of up to 93% without an assistant in-plane field. The spin-polarized electrons, aligned along the y-direction, undergo spin reflection at the NiO interface featuring noncollinear spin structures with canted magnetization, thereby generating an out-of-plane spin polarization component $\sigma_z$. Our thorough validation of the $\sigma_z$ contribution, as evidenced by measurements of the shift effect in out-of-plane hysteresis loops, underscores the pivotal role of this mechanism in achieving field-free SOT switching. Simultaneously, we observed that the ratio of field-free SOT switching gradually decreases with the increasing thickness of the NiO layer. Beyond advancing our fundamental understanding of spintronics, this breakthrough opens promising avenues for practical applications, marking a significant stride toward realizing innovative technologies in the field.

## SUPPLEMENTARY MATERIAL

Refer to the supplementary material for additional information.

## ACKNOWLEDGEMENTS

This work is supported by the

The authors would like to thank the staff from BL07U beamline of the Shanghai Synchrotron Radiation Facility (SSRF) for assistance with XMCD/XAS data collection.

## DATA AVAILABILITY

The data that support the findings of this work are available from the corresponding author upon reasonable request.